\documentclass[prb,8pt,twocolumn,showpacs,floats]{revtex4}
\usepackage{amssymb,graphicx,color,amsmath}
\usepackage[all]{xy}

\renewcommand{\Im}{{\mathrm{Im}}}
\renewcommand{\Re}{{\mathrm{Re}}}
\newcommand{\sign}{{\mathrm{sign}}}

\SelectTips{eu}{12}
\newcommand{\x}{\object@{>}}
\newcommand{\y}{\object@{<}}

\newcommand{\av}[1]{\ensuremath{\left\langle{#1}\right\rangle}}
\newcommand{\ns}{\ensuremath{\!\!}}

\newcommand{\verfull}[5]{{\vcenter{\xymatrix @-0.6pc @R=1.4pc  @M=0pt { 
\ar@{-}[r]|{\x}^{#1}&\ar@{-}[d]\ar@{}[dr]|{#3}\ar@{-}[r]&\ar@{-}[d]  \ar@{-}[r]|{\x}^{#4}&\\  
 &\ar@{-}[l]|{\x}^{#2} \ar@{-}[r] &  &  \ar@{-}[l]|{\x}^{#5} \\ }}}}

\newcommand{\verfulldots}[5]{{\vcenter{\xymatrix @-0.6pc @R=1.4pc  @M=0pt { 
\ar@{-}[r]|{\x}="a"^{#1}&\ar@{-}[d]\ar@{}[dr]|{#3}\ar@{-}[r]&\ar@{-}[d]  \ar@{-}[r]|{\x}="c"^{#4}&\\  
 &\ar@{-}[l]|{\x}="b"^{#2} \ar@{-}[r] &  &  \ar@{-}[l]|{\x}="d"^{#5} 
\ar@{} "a";"b" |*+={...} \ar@{} "c";"d" |*+={...} \\ }}}}

\newcommand{\verpair}[2]{ \vcenter{\xymatrix @-0.6pc @R=1.4pc  @M=0pt{  
\ar@{-}[r]|{\x}^{#1} &\\  &  \ar@{-}[l]|{\x}^{#2} \\ }}}

\newcommand{\verpairdots}[2]{ \vcenter{\xymatrix @-0.6pc @R=1.4pc  @M=0pt{  
\ar@{-}[r]|{\x}="a"^{#1} &\\  &  \ar@{-}[l]|{\x}="b"^{#2} \ar@{} "a";"b" |*+={...}  }}}

\newcommand{\verbare}[5]{ \vcenter{\xymatrix @-0.6pc @R=1.4pc  @M=0pt{  
\ar@{-}[r]|{\x}^{#1} &\ar@{~}[d]^{#3} \ar@{-}[r]|{\x}^<(.8){#4} &\\  &
\ar@{-}[l]|{\x}^{#2} & \ar@{-}[l]|{\x}^{#5} \\ }}}

\newcommand{\verbaretot}[5]{  \vcenter{\xymatrix @-0.6pc @R=1.4pc  @M=0pt{  
\ar@{-}[r]|{\x}^{#1} &\ar@{-}@<-0.2pc>[d] \ar@{-}@<0.2pc>[d]^{#3} \ar@{-}[r]|{\x}^<(.8){#4}& \\ &\ar@{-}[l]|{\x}^{#2} & \ar@{-}[l]|{\x}^{#5}  }}}

\newcommand{\verbarereg}[5]{  \vcenter{\xymatrix @-0.6pc @R=1.4pc  @M=0pt{  
\ar@{-}[r]|{\x}^{#1} &
\ar@{-}@<-0.3pc>[d]|<(0.1){.}="a"|<(0.35){.}="b"|<(0.6){.}="c" 
\ar@{-}@<0.3pc> [d]|<(0.4){.}="d"|<(0.65){.}="e"|<(0.9){.}="f" 
\ar@{}[d]^{#3} \ar@{-}[r]|{\x}^<(.8){#4}&
\ar@{-} "a";"d" \ar@{-} "b";"e" \ar@{-} "c";"f" 
\\&\ar@{-}[l]|{\x}^{#2} & \ar@{-}[l]|{\x}^{#5}  }}}

\newcommand{\vertexone}[5]{  \vcenter{\xymatrix @-0.6pc @R=1.4pc  @M=0pt{  
\ar@{-}[r]|{\x}="x"^{#1} &
\ar@{-}@<-0.3pc>[d]|<(0.1){.}="a"|<(0.35){.}="b"|<(0.6){.}="c" 
\ar@{-}@<0.3pc> [d]|<(0.4){.}="d"|<(0.65){.}="e"|<(0.9){.}="f" 
\ar@{}[d]^{#3} \ar@{-}[r]|{\x}="z"^<(.8){#4}&
\ar@{-} "a";"d" \ar@{-} "b";"e" \ar@{-} "c";"f" 
\\ & \ar@{-}[l]|{\x}="y"^{#2} & \ar@{-}[l]|{\x}="t"^{#5} 
\ar@{} "x";"y" |*+={...} \ar@{} "z";"t" |*+={...}   }}}

\newcommand{\vertextwo}{  \vcenter{\xymatrix @-0.6pc @R=1.4pc  @M=0pt{  
\ar@{-}[r]|{\x}="x" &
\ar@{-}@<-0.3pc>[d]|<(0.1){.}="a"|<(0.35){.}="b"|<(0.6){.}="c" 
\ar@{-}@<0.3pc> [d]|<(0.4){.}="d"|<(0.65){.}="e"|<(0.9){.}="f" 
\ar@{-}[r]|{\x}="xx" & 
\ar@{-}@<-0.3pc>[d]|<(0.1){.}="aa"|<(0.35){.}="bb"|<(0.6){.}="cc" 
\ar@{-}@<0.3pc> [d]|<(0.4){.}="dd"|<(0.65){.}="ee"|<(0.9){.}="ff" 
\ar@{-}[r]|{\x}="xxx" & 
\\ & \ar@{-}[l]|{\x}="y" & \ar@{-}[l]|{\x}="yy" & \ar@{-}[l]|{\x}="yyy" 
\ar@{-} "a";"d" \ar@{-} "b";"e"  \ar@{-} "c";"f" \ar@{-} "aa";"dd" 
\ar@{-} "bb";"ee" \ar@{-} "cc";"ff" 
\ar@{} "x";"y" |*+={...} \ar@{} "xx";"yy" |*+={...} \ar@{} "xxx";"yyy" |*+={...} }}}

\newcommand{\versteptot}[8]{  \vcenter{\xymatrix  @-0.6pc @R=1.4pc  @M=0pt{
\ar@{-}[r]|{\x}^{#1}&\ar@{-}[d]\ar@{}[dr]|{#3}\ar@{-}[r] & \ar@{-}[d]  
\ar@{-}[r]|{\x}^{#4}& \ar@{-}@<-0.2pc>[d] \ar@{-}@<0.2pc>[d]^{#6} 
\ar@{-}[r]|{\x}^<(.8){#7} & 
\\ &\ar@{-}[l]|{\x}^{#2} \ar@{-}[r] &  &  \ar@{-}[l]|{\x}^{#5}  &  \ar@{-}[l]|{\x}^{#8}  }}}

\newcommand{\verstep}[8]{  \vcenter{\xymatrix  @-0.6pc @R=1.4pc  @M=0pt{
\ar@{-}[r]|{\x}^{#1}&\ar@{-}[d]\ar@{}[dr]|{#3}\ar@{-}[r] & \ar@{-}[d]  
\ar@{-}[r]|{\x}^{#4}&\ar@{~}[d]^{#6}\ar@{-}[r]|{\x}^<(.8){#7} & \\ 
 &\ar@{-}[l]|{\x}^{#2} \ar@{-}[r] &  &  \ar@{-}[l]|{\x}^{#5}  &  \ar@{-}[l]|{\x}^{#8}  }}}

\newcommand{\verstepdots}[8]{ \vcenter{\xymatrix  @-0.6pc @R=1.4pc  @M=0pt{
\ar@{-}[r]|{\x}^{#1}&\ar@{-}[d]\ar@{}[dr]|{#3}\ar@{-}[r] & \ar@{-}[d]  
\ar@{-}[r]|{\x}="a"^{#4}&\ar@{~}[d]^{#6}\ar@{-}[r]|{\x}^<(.8){#7} & \\ 
 &\ar@{-}[l]|{\x}^{#2} \ar@{-}[r] &  &  \ar@{-}[l]|{\x}="b"^{#5}  &  \ar@{-}[l]|{\x}^{#8}  \ar@{} "a";"b" |*+={...} }}}

\newcommand{\tadpole}[4]{{{ \xymatrix @=1pc @M=0pt @C=1pc{
 \ar@{-}[r]|{\x}_{#1}  & \ar@/^1.1pc/@{~}[rr]^{#2} \ar@{-}[rr]|{\x}_{#3} &&  \ar@{-}[r]|{\x}_{#4} &  }}}} 

\newcommand{\lefttriang}{{\vcenter{ \xymatrix  @-1.5pc @R=.8pc  @M=0pt{ 
& \ar@{-}[dd] \\ \ar@{-}[ur] \ar@{-}[dr] & \\ & }}}}

\newcommand{\righttriang}{{\vcenter{\xymatrix  @-1.5pc @R=.8pc  @M=0pt{  
\ar@{-}[dd] & \\ & \ar@{-}[ul] \ar@{-}[dl] \\ & }}}}

\begin{document}

\title{Long Wavelength Correlations and Transport in a Marginal Fermi Liquid.}
\author{A.~Shekhter, and C.~M.~Varma}
\affiliation{Department of Physics and Astronomy, University of California, Riverside, California 92521}
\begin{abstract}
Marginal Fermi liquid was originally introduced as a phenomenological description of the cuprates in a part of the metallic doping range which appears to be governed by fluctuations due to a quantum-critical point. An essential result due to the form of the assumed fluctuation spectra is that the large inelastic quasiparticle relaxation rate near the Fermi-surface is proportional to the energy measured from the chemical potential, $\tau_i^{-1}\propto\varepsilon$.  We present a  microscopic long-wavelength derivation of the hydrodynamic properties in such a situation by an extension of the procedure that Eliashberg used for the derivation of the hydrodynamic properties of a Landau-Fermi-liquid. In particular, the density-density and the current-current correlations and the relation between the two are  derived, and the connection to microscopic calculations of the frequency dependence of the optical conductivity with an additional Fermi-liquid correction factor shown to follow.  The method used here may be necessary, quite generally, for the correct hydrodynamic theory for any problem of quantum-critical fluctuations in fermions.
\end{abstract}
\date{\today}
\maketitle

\section{Introduction}

It is necessary for the success of the conventional Fermi liquid description \cite{Pitaevskii}  that the decay rate for quasiparticles close to Fermi surface is small compared to the energy of the quasiparticle,   $\tau^{-1}\ll\varepsilon$.  This is followed in conventional metals where the  inelastic collision rate for quasiparticles near the Fermi surface due to electron-electron interactions leads to an energy relaxation via creation of particle-hole pair with a rate $\tau_i^{-1}\propto\varepsilon^2/\varepsilon_F$.  On the one hand, this ensures that Landau quasiparticles are well defined close to the Fermi surface, on the other, in this situation transport phenomena are dominated by other sources of scattering, such as lattice disorder. But in the original Landau Fermi liquid system, liquid He$^3$, there is no other scattering process to mask the He-He atom collision rate in the very low frequency limit. As pointed out by Eliashberg \cite{Eliashberg}, the microscopic derivation of the Fermi liquid state has to be modified from the simpler procedure used outside this regime, if static or ultra-low frequency or hydrodynamic $\omega\lesssim\tau_i^{-1}$ phenomena are to be described correctly. For liquid He$^3$ such modification was devised by Eliashberg for analysis of the first sound in He$^3$ which, by its nature, is an ultra-low frequency phenomenon.

The metallic state of the Copper-Oxide in a funnel shaped region emanating from a point at doping $x=x_c$ at $T=0$ has unusual transport properties which appear to be governed by fluctuations due to a quantum critical point. A phenomenological spectral function for the fluctuations was given in Ref.~\onlinecite{Varma} which leads to the concept of the marginal Fermi-liquid (MFL). These fluctuations have a scale-invariant singular low energy form but a smooth momentum-dependence as for spatially local fluctuations. In the marginal Fermi-liquid (MFL), the decay rate of the single-particle excitations and their energy and momentum relaxation rate are all $\propto \varepsilon$, the energy of the excitation itself. The theory explained the observed unusual  temperature and frequency dependence of the transport properties such as resistivity and optical conductivity and Hall effect \cite{HallVarmaAbrahams}. Predictions of the phenomenology for the fluctuations moreover were satisfied in the long wavelength limit in Raman scattering experiments and for the single-particle spectra in ARPES experiments \cite{Valla,Kaminski,ZhuVarma,RamanExp}. 

The relaxation rates $\propto\varepsilon$ imply that in copper-oxide metals in the quantum-critical or MFL regime,  every hydrodynamic transport phenomena is in the "ultra-low" frequency regime and a derivation of the hydrodynamic properties similar to the one designed by Eliashberg for the Fermi-liquid is required. This is necessary in order to formally prove the validity of the simple calculations of such properties \cite{Varma,Abrahams1996} which agree with experiments.  It should be noted that a hydrodynamic form for  density-density correlation function in a MFL obeying the continuity equation was also suggested \cite{Kotliar1991}. It is important to show that this form is not an independent assumption but follows from the original assumption of a local quantum-critical spectra. In this paper, we follow an extension of the Eliashberg theory to achieve these goals. The locality in the quantum-critical fluctuation spectra of the cuprates makes such an extension relatively easy.  The consistent derivation of the hydrodynamic properties in more general quantum-critical problems \cite{VNW} may require further developments of the methods used here.

The results of this paper are independent of the microscopic underpinnings of the quantum-critical point and the quantum critical fluctuations and their coupling to fermions in the cuprates. For the interested reader, we mention that the fluctuations have been recently derived as the fluctuations (of the flux variables)\cite{AjiVarma}  due to the quantum critical point of a loop-current order predicted \cite{Varma1997} and by now observed in three different families of cuprates \cite{NeutronExp}. The coupling-function of the fluctuation to the fermions has also been determined \cite{ASV}.  
 
\section{Particle-hole ladder in marginal Fermi liquid.}
 
We wish to describe the transport phenomena in MFL system  consistently in terms of the particle-hole ladder in a manner similar to Fermi liquid description of conventional metal. Particle-hole ladder describes multiple rescattering of the particle-hole pairs and is the essential mathematical object in the two-particle correlation functions (density-density, current-current, etc.) which determine transport properties of fermions. It has a singular dependence on the order of limits $\omega,q\rightarrow0$ which is dictated by the conservation laws. To be specific, we consider the density-density correlation function $\av{\rho\rho}_{\omega,q}$. Particle number conservation requires that density correlation function vanishes in the limit  $\lim\limits_{q\rightarrow0}\av{\rho\rho}_{\omega,q}=0$. In the opposite limit it equals the compressibility  $\lim\limits_{\omega\rightarrow0}\av{\rho\rho}_{\omega,q}=dn/d\mu$ and is finite. Direct generalization of the microscopic Fermi liquid calculation to account for inelastic processes fails to produce a function that meets these constraints at $\omega,qv_F <\tau_i^{-1}$ because dynamic nature of the interaction amplitudes is ignored. We now discuss in detail the technical steps to show that conservation laws are indeed obeyed by a  MFL. 

It is necessary for us to specify only that the  quantum-critical fluctuations in a MFL are described by a correlation function\cite{AjiVarma}
\begin{align}\label{eq:local}
	\Im\chi^R(q,\omega) =&\left\{		\begin{matrix} 
	-\chi_0 \sign\omega,  &|\omega| \ll \omega_0 \\
	 0, & |\omega| \gg\omega_0 		\end{matrix}\right.	\notag\\
	\Re\chi(\omega) =& -\chi_0 \frac2{\pi} \ln\frac{\omega_0}{|\omega|}\,.
\end{align}
In cuprates the upper cutoff at $\omega_0$ appears to be sharp enough to have observable consequences, see Ref~\onlinecite{ZhuVarma}.
 In this paper we will discuss only properties at $T=0$. At zero temperature the spectral weight $\Im\chi^R(q,\omega)$ of the local fluctuations extends all the way from the upper cutoff $\omega_0$ to the zero energy. In cuprates the upper cut-off $\omega_0\approx0.5eV$ is few times smaller than $\varepsilon_F$. The inelastic decay rate of fermions is controlled by emission/absorption of such local fluctuation. The fact that the mode is local, i.e., its correlation function $\chi(\omega,\mathrm{q})$ independent of $\mathbf{q}$,  facilitates our analysis. At zero temperature the quasiparticle self-energy is given by
\begin{align}\label{eq:selfenergy}
&\Im\Sigma^R(\varepsilon)= \tadpole{}{}{}{} \;=\;
\gamma^2\nu\int_0^{\varepsilon}\!\!\!\!d\omega \;\Im\chi^R(\omega) \notag\\
&\quad= -g \left\{\begin{matrix} |\varepsilon|, &\qquad |\varepsilon| \ll \omega_0 \\ \omega_0, &\qquad |\varepsilon| \gg \omega_0 \end{matrix}\right. \,.
\end{align} 
Here $\gamma$ is a coupling constant for the interaction between fermions and the fluctuations.\cite{rotor}  $\nu$ is the density of states for one spin species. We have  introduced dimensionless coupling constant $g=\nu\chi_0\gamma^2$; in cuprates  $g\lesssim1$. Owing to the fact that the upper energy cutoff $\omega_0$ of the local fluctuations is smaller than Fermi energy, this calculation is in fact self consistent, i.e., the same answer to order $\omega_0/E_F$ is obtained if one uses full Green's function $G^R(\varepsilon,p)=1/[\varepsilon-\epsilon(p)+\mu-\Sigma^R(\varepsilon)]$ in the intermediate section. The real part $\Re\Sigma^R(\varepsilon)$ follows from Kramers-Kronig relation. For $\varepsilon\ll\omega_0$ (this is enough for our purposes as we consider physics close to the Fermi surface )  we find~:
\begin{align}\label{eq:selfenergy-real}
	&\Re \Sigma^R(\varepsilon\ll\omega_0) = 
	- g \frac2{\pi} \left( \ln \frac{\omega_0}{|\varepsilon|} + 1 \right) \varepsilon \,. 
\end{align}
Behavior of $\Sigma(\varepsilon)$ at and around $\varepsilon\approx\omega_0$ has a direct experimental consequence \cite{Valla} as is discussed in Ref.~\onlinecite{ZhuVarma}. As a result of singular form of the self-energy, the single-particle Green's function does not have a simple pole anymore.

In calculation of two-particle correlation functions within microscopic Fermi-liquid formalism one distinguishes dynamic and static part, $\av{\rho\rho}_{\omega,q}^{\text{tot}}=\av{\rho\rho}_{\omega,q}^{\text{dyn}}+\av{\rho\rho}_{\omega,q}^{\text{stat}}$. The static part, $\av{\rho\rho}_{\omega,q}^{\text{stat}}=\lim\limits_{\omega/q\rightarrow0}\av{\rho\rho}_{\omega,q}^{\text{tot}}$ does not have significant dependence on $\omega,q$ near the Fermi surface,  $\av{\rho\rho}_{\omega,q}^{\text{stat}}=\lim\limits_{\omega\rightarrow0}\av{\rho\rho}_{\omega,q}^{\text{tot}}=dn/d\mu$. The singular behavior of $\av{\rho\rho}_{\omega,q}^{\text{tot}}$ is inherited by the dynamic part $\av{\rho\rho}_{\omega,q}^{\text{dyn}}$. Indeed, from its definition, $\av{\rho\rho}_{\omega,q}^{\text{dyn}}$ vanishes in the limit $\omega/q\rightarrow0$, i.e.,  $\lim\limits_{\omega\rightarrow0}\av{\rho\rho}_{\omega,q}^{\text{dyn}}=0$; in the opposite limit it is finite $\lim\limits_{q\rightarrow0}\av{\rho\rho}_{\omega,q}^{\text{dyn}}=-dn/d\mu$. Physical motivation behind separation into dynamic and static parts is that $\av{\rho\rho}_{\omega,q}^{\text{dyn}}$ gets its contribution from excitations close to the Fermi surface and can be analyzed in terms of quasiparticles. For correlation functions of conserved quantities the static part can be determined from the Ward identity, see, e.g.,  Refs.~\onlinecite{Pitaevskii},~\onlinecite{Leggett,Baym}.

We use the temperature technique\cite{Pitaevskii} to calculate the density correlation function. 
We replace summation over frequency in Eq.~(\ref{eq:betheall}) with integrals on the real axis \cite{Eliashberg}. General Matsubara correlation functions consist of several analytic pieces connected along  branch cuts in the complex plane of its energy arguments.  A discontinuity across the branch cut has the  direct physical meaning of a spectral weight related to the particular physical excitation channel. For instance, the single-particle Green's function consists of two analytic pieces, $G^R(\varepsilon)$ and $G^A(\varepsilon)$, connected along the real axis, $\Im\varepsilon=0$ in the complex-$\varepsilon$ plane; the spectral weight $G^R(\varepsilon)-G^A(\varepsilon)=2\Im G^R(\varepsilon)$ describes fermion quasiparticles. Analytic structure of $\chi(\omega)$ is similar. The particle-hole ladder $\Gamma(\varepsilon_1,\varepsilon_2;\omega)$, a function of three energy variables can similarly be splitted into several analytic pieces connected along branch cuts. Within this formalism understanding  analytic structure of $\Gamma(\varepsilon_1,\varepsilon_2;\omega)$ is crucial in capturing the singular contribution to the particle-hole ladder. 

The dynamic part of the density correlation function is given by 
\begin{align}\label{eq:densitytot}
&\av{\rho\rho}_{\omega,q}^{\text{dyn}}  =  \lefttriang\verpair{i\varepsilon+i\omega}{i\varepsilon}\righttriang  + \lefttriang\verfull{i\varepsilon_1+i\omega}{i\varepsilon_1}{\Gamma}{i\varepsilon_2+i\omega}{i\varepsilon_2}\righttriang
\end{align}
where vertex $\Gamma(\varepsilon_1,\varepsilon_2;\omega)$ (the particle-hole ladder) describes multiple rescattering of particle-hole pair~:
\begin{align}\label{eq:betheall}
  \verfull{i\varepsilon_1\!+\!i\omega}{i \varepsilon_1}{\Gamma}{i\varepsilon_2\!+\!i\omega}{i \varepsilon_2}
  \!\!\!=\!\!\!\verbaretot{i\varepsilon_1\!+\!i\omega}{i \varepsilon_1}{\,}{i\varepsilon_2\!+\!i\omega}{i\varepsilon_2}
\!\!\!+\!\!\! \versteptot{i\varepsilon_1\!+\!i\omega}{i\varepsilon_1}{\Gamma}{i\varepsilon\!+\!i\omega}{i\varepsilon}{\,}{i\varepsilon_2\!+\!i\omega}{i\varepsilon_2} \,.
\end{align}
Here a pair of horizontal lines describes a free propagation of particle-hole pair. The rescattering is induced by the elementary vertex represented by narrow rectangle. It contains all processes that cannot be separated in two independent pieces by cutting a particle and a hole line. 

The retarded correlation function $\av{\rho\rho}_{\omega,q}^{\text{dyn}}$ is determined by analytic continuation from upper half of the complex   $\omega$-plane, $\omega_n>0$. Mathematical representation of the fact that $\av{\rho\rho}_{\omega,q}^{\text{dyn}}$ gets its contribution from quasiparticles near the Fermi surface is that particle-hole sections in Eq.~(\ref{eq:densitytot}) are restricted to the region  $\varepsilon_n<0$ and $\varepsilon_n+\omega_n>0$. The analytic continuation of two Green's functions in the  particle-hole section (any part in the  diagram where it can be cut into two by cutting a particle and a hole line) from the region  $\varepsilon_n<0$ and $\varepsilon_n+\omega_n>0$ to the real axis defines the dynamic  particle-hole section (RA section) in which the upper fermion line is $G^R$ and the lower line is $G^A$. We define function $S^R$ by
\begin{align} \label{eq:RAsection}
&2\pi i\nu S^R(\varepsilon; \omega,q) =   \verpair{i\varepsilon+i\omega}{i\varepsilon}  = \int\frac{d^2p}{(2\pi)^2} G_{\mathbf{p+q}}^R(\varepsilon+\omega) G_{\mathbf{p}}^A(\varepsilon) \notag\\
& = 2\pi i\nu\int\frac{d\theta}{2\pi} \frac1{\omega-qv_F\cos\theta-[\Sigma^R(\varepsilon+\omega)-\Sigma^A(\varepsilon)]}
\end{align} 
where the second line is valid in the limit $qv_F\ll\varepsilon_F$. In our analysis of the ladder  [Eq.~(\ref{eq:betheall})] we choose to describe the elementary vertex in terms of the amplitude $\Gamma^k$, see Ref.~\onlinecite{Pitaevskii}; in this case  all intermediate particle-hole sections are RA sections given by Eq.~(\ref{eq:RAsection}).

In MFL system local fluctuations make a singular contribution to the amplitude $\Gamma^k$. In order to single out the singular part of the particle-hole ladder we separate the singular and non-singular parts of the elementary vertex (the thin rectangle in Eq.~(\ref{eq:betheall}))~:
\begin{align}\label{eq:elementaryvertex}
&\verbaretot{}{}{}{}{} = \verbare{}{}{}{}{} + \verbarereg{}{}{}{}{} \,.
\end{align}
The non-singular contributions to the amplitude $\Gamma^k$, the second term in this equation,  account for all processes that do not lead to sharp dependences near the Fermi surface; these include non-singular exchange by local fluctuation, such as
\begin{align}\label{eq:nonsingular}
&\verbarereg{}{}{}{}{} = \vcenter{\xymatrix  @C=1pc @R=1.4pc  @M=0pt{
\ar@{-}[r]|{\x} & \ar@{~}[dr]\ar@{-}[d]|{\x} & \ar@{-}[r]|{\x}   &    \\ 
&\ar@{-}[l]|{\x}  \ar@{~}[ur]  &  \ar@{-}[u]|{\x}  &  \ar@{-}[l]|{\x} \\ }} 
+ 
\vcenter{\xymatrix   @C=1pc @R=1.4pc  @M=0pt{
\ar@{-}[r]|{\x} & \ar@{~}[dr]\ar@{-}[r]|{\x} & \ar@{-}[r]|{\x}  \ar@{~}[dl] &    \\ 
&\ar@{-}[l]|{\x}  &  \ar@{-}[l]|{\x} &  \ar@{-}[l]|{\x} \\ }}
+
\vcenter{\xymatrix   @C=1pc @R=1.4pc  @M=0pt{
\ar@{-}[r]|{\x}& \ar@{-}[r]|{\x} \ar@{~}@/^/[rr] & \ar@{~}[d]\ar@{-}[r]|{\x}  &  \ar@{-}[r]|{\x}  &   \\ 
& & \ar@{-}[ll]|{\x}  & &  \ar@{-}[ll]|{\x} \\ }}
+ \cdots
\end{align}
as well as direct electron-electron interaction processes. In the zero harmonic spin-symmetric channel we describe the sum of all such contributions to $\Gamma^k$ by Landau parameter  $B_0 =\nu\langle\Gamma^k\rangle$; here averaging $\langle\cdots\rangle$ is over the Fermi surface to give zero harmonic. In phenomenological calculations one typically uses the amplitude $\Gamma^{\omega}$ which in the zero harmonic spin-singlet channel is determined by Landau parameter $F_0^s$. The parameter $B_0$ is related to $F_0^s$ through the relation $1+B_0=1/(1+F_0^s)$. The role of the non-singular part of the elementary vertex is to introduce multiplicative renormalizations which are typical for any Fermi liquid system. Let us comment on why processes in Eq.~(\ref{eq:nonsingular}) do not lead to singular contributions in $\Gamma^k$. The feature they all share is presence of additional frequency integration which is not restricted to the vicinity of the Fermi surface. Such additional integration smears the discontinuity of the correlation function of the local fluctuation. In addition, the logarithmic factor in Eq.~(\ref{eq:local}) is not effective here because due to additional integration the typical frequency carried by local fluctuation is not small. 

The singular contribution to the elementary vertex is given by the first term in Eq.~(\ref{eq:elementaryvertex}). It represents a particle-hole rescattering via exchange of a single local fluctuation. In view of Eq.~(\ref{eq:elementaryvertex}) we make a partial resummation in  Eq.~(\ref{eq:densitytot}) 
\begin{align}\label{eq:density}
&\av{\rho\rho}_{\omega,q}^{\text{dyn}}  =  \lefttriang\Bigg[
\verpairdots{i\varepsilon+i\omega}{i\varepsilon} 
+\vertexone{}{}{}{}{}
+\vertextwo+\cdots
\Bigg]\righttriang
\end{align}
The particle hole ladder breaks into a sequence of singular segments (dotted sections in this equation) connected by non-singular amplitude $B_0$, 
\begin{align} \label{eq:dressedsection}
&\verpairdots{}{}  = \verpair{}{} + \verfull{}{}{\Gamma_s}{}{}\,.
\end{align}
Equation~(\ref{eq:dressedsection}) defines the singular  vertex  $\Gamma_s(\varepsilon_1,\varepsilon_2;\omega)$ which is a result of multiple rescattering with singular elementary vertex only~:
\begin{align} \label{eq:bethe}
&\verfull{i\varepsilon_1\!+\!i\omega}{i\varepsilon_1}{\Gamma_s}{i\varepsilon_2\!+\!i\omega}{i\varepsilon_2}
\!\!\!=\!\!\!\verbare{i\varepsilon_1\!+\!i\omega}{i\varepsilon_1}{\,}{i\varepsilon_2\!+\!i\omega}{i\varepsilon_2}
\!\!\!+\!\!\! \verstep{i\varepsilon_1\!+\!i\omega}{i\varepsilon_1}{\Gamma_s}{i\varepsilon\!+\!i\omega}{i\varepsilon}{\,}{i\varepsilon_2\!+\!i\omega}{i\varepsilon_2} \,.
\end{align}
Frequency sum in the successive dotted sections in Eq.~(\ref{eq:density}) are decoupled since they are separated by nonsingular amplitude $B_0$. In contrast, the frequency summations in successive sections in ladder equation for singular vertex, Eq.~(\ref{eq:bethe}), are coupled because the local fluctuations introduce essential dependence on $\varepsilon_1-\varepsilon_2$. Here the singular behavior of $\chi(\omega)$, see Eq.~(\ref{eq:local}), translates into essential dependence on $\varepsilon_1,\varepsilon_2$. We emphasize that singular dependence of $\Gamma_s$ on $\varepsilon_1,\varepsilon_2$ is purely dynamic, i.e., it is not associated with or results from any peculiar momentum dependence. Equation~(\ref{eq:bethe}) completes identification of the singular vertex in the particle-hole ladder. To analyze Eq.~(\ref{eq:bethe}) we have to perform analytic continuation to the real axis from the frequency interval $i\varepsilon_{1,2}<0$ and $i\varepsilon_{1,2}+i\omega>0$. This requires analysis of the analytic structure of the vertex $\Gamma_s(\varepsilon_1,\varepsilon_2;\omega)$.

In the complex plane of two fermion frequencies, $\varepsilon_1,\varepsilon_2$, the analytic structure of the vertex  $\Gamma_s(\varepsilon_1,\varepsilon_2;\omega)$ can be illustrated by a diagram\cite{Eliashberg}~:
\begin{align}\label{eq:eliashberg}
\vcenter{\xymatrix @M=0pt @C-0.5pc @R-0.5pc @-0.3pc{
&&&&&\\ &&&&&\\
\ar@{}[r]_(0.5)*-{-\omega} \ar[rrrr] \ar@{}[rrrrr]_(0.95)*+{\Im\varepsilon_1} & \ar@{}[rd]|*+={I} &&&& \\
&&\ar@{}[rd]|*+={II}&&&\\  \ar@{-}[rrrr] &&&&& \\
\ar@{-}[uuuurrrr]&\ar@{-}[uuuu]&&\ar@{}[u]_(0.7)*-{-\omega} \ar[uuuu] & \ar@{}[uuuuu]^(0.9)*+{\Im\varepsilon_2} &}}
\end{align}
Each line represents a branch cut connecting different analytic pieces.\cite{anotherBC} The horizontal and vertical lines indicate the branch cuts that are already present in the external lines, $\Im\varepsilon_{1,2}=0$ and $\Im\varepsilon_{1,2}+\omega=0$.  The diagonal line in Eq.~(\ref{eq:eliashberg}), represents the discontinuity at $\Im(\varepsilon_1-\varepsilon_2)=0$ in $\Gamma_s(\varepsilon_1,\varepsilon_2;\omega)$ (and the corresponding singularity in the interaction amplitude $\Gamma^k$); it  is induced by the discontinuity in $\chi(\omega)$ at $\Im\omega=0$. In the mathematical language of temperature technique the branch cut at $\Im\varepsilon_1=\Im\varepsilon_2$ (and associated singularity) represents the physical effect of multiple rescattering via the exchange of local fluctuations.

The  analytic continuation of $\Gamma_s(\varepsilon_1,\varepsilon_2;\omega)$ to the real axis $\Im\varepsilon_{1,2}=0$ from the frequency interval $i\varepsilon_{1,2}<0$ and $i\varepsilon_{1,2}+i\omega>0$ is not unique as a result of the branch cut at $\Im\varepsilon_1-\Im\varepsilon_2=0$. On the real axis the singular vertex $\Gamma_s$ is determined by two functions $\Gamma^{I,\;II}$ depending on the order in which $\Im\varepsilon_{1,2}$ are sent to zero~: 
\begin{align}
	\verfull{}{}{\Gamma^{II}}{}{} = \Gamma_s(\Im\varepsilon_1>\Im\varepsilon_2; \; \Im\varepsilon_{1,2}\rightarrow-0) \,, \notag\\
	\verfull{}{}{\Gamma^{I}}{}{} = \Gamma_s(\Im\varepsilon_1<\Im\varepsilon_2; \; \Im\varepsilon_{1,2}\rightarrow-0) \,.
\end{align}
Making analytic continuation to the real axis in Eq.~(\ref{eq:bethe}) we obtain coupled equations satisfied by  functions $\Gamma^{II,\;I}$~:
\begin{align}\label{eq:realaxis}
&\verfull{\varepsilon_1+\omega}{\varepsilon_1}{\Gamma^{II}}{\varepsilon_2+\omega}{\varepsilon_2} = 
\verbare{}{}{R}{}{} - \int\frac{d\varepsilon}2 \times \notag\\
&\Big[- {\scriptstyle\tanh\frac{\varepsilon}{2T}}\verstep{}{}{I}{\varepsilon+\omega}{\varepsilon}{R}{}{} 
+ {\scriptstyle\coth\frac{\varepsilon-\varepsilon_1}{2T}} \verstep{}{}{I\!-\!{II}}{}{}{R}{}{} 
\notag\\
&+ {\scriptstyle\coth\frac{\varepsilon-\varepsilon_2}{2T}}\!\verstep{}{}{{II}}{}{}{R-A}{}{} 
+ {\scriptstyle\tanh\frac{\varepsilon+\omega}{2T}}\!\verstep{}{}{{II}}{}{}{A}{}{} 
\;\;\Big] \,,
\notag\\
&\verfull{\varepsilon_1+\omega}{\varepsilon_1}{\Gamma^I}{\varepsilon_2+\omega}{\varepsilon_2} = 
\verbare{}{}{A}{}{}- \int\frac{d\varepsilon}2  \times\notag\\
&\Big[- {\scriptstyle\tanh\frac{\varepsilon}{2T}}\!\verstep{}{}{I}{\varepsilon+\omega}{\varepsilon}{R}{}{} 
+ {\scriptstyle\coth\frac{\varepsilon-\varepsilon_1}{2T}}\!\verstep{}{}{I\!-\!{II}}{}{}{A}{}{} 
\notag\\
&+ {\scriptstyle\coth\frac{\varepsilon-\varepsilon_2}{2T}}\!\verstep{}{}{I}{}{}{R-A}{}{} 
+ {\scriptstyle\tanh\frac{\varepsilon+\omega}{2T}}\!\verstep{}{}{{II}}{}{}{A}{}{} 
\;\;\Big]
\end{align} 
Each line or vertex in this equation is assigned a particular analytic piece continued to the real axis as indicated; frequency and momenta are assigned here in the same way as in Eq.~(\ref{eq:bethe}) with subsequent replacement $i\varepsilon\rightarrow\varepsilon$. In all particle-hole sections, upper and lower fermion lines are $G^R$ and  $G^{A}$ respectively. To proceed we first have to make sure that frequency integration is restricted to finite interval; only then we are allowed to perform a momentum integration in the intermediate section, see Eq.~(\ref{eq:RAsection}). This is achieved by rewriting Eq.~(\ref{eq:realaxis}) in terms of   $\Gamma^{\pm}=\nu(\Gamma^{II}\pm\Gamma^{I})/2$. Indeed, all hyperbolic functions will enter in the combinations such as  $\int d\varepsilon[\tanh\frac{\varepsilon+\omega}2 - \tanh\frac{\varepsilon}2]/2$ etc. In particular, at zero temperature we use the following identities 
\begin{align}\begin{aligned}
&{\scriptstyle\int\frac{d\varepsilon}2 [\tanh\frac{\varepsilon+\omega}{2T} - \tanh\frac{\varepsilon}{2T}] \cdots }
&\rightarrow&{\scriptstyle\int\limits_{-\omega}^0 {d\varepsilon} \cdots} \\
&{\scriptstyle\int\frac{d\varepsilon}2 [ -\tanh\frac{\varepsilon+\omega}{2T} - \tanh\frac{\varepsilon}{2T} + 2\coth\frac{\varepsilon-\varepsilon_2}{2T}] \cdots} 
&\rightarrow&{\scriptstyle\int\limits_{-\omega}^0 {d\varepsilon}\, \sign(\varepsilon-\varepsilon_2) \cdots} \\
&{\scriptstyle\int \frac{d\varepsilon}2[\coth\frac{\varepsilon-\varepsilon_2}{2T}-\coth\frac{\varepsilon-\varepsilon_1}{2T}] \cdots }
&\rightarrow&{\scriptstyle\sign(\varepsilon_1-\varepsilon_2)\int\limits_{\min\varepsilon_1,\varepsilon_2}^{\max\varepsilon_1,\varepsilon_2} d\varepsilon \cdots }
\end{aligned}\notag\end{align}
where both $\varepsilon_1,\varepsilon_2$ are assumed to belong to the interval $[-\omega,0]$. Combining equations Eq.~(\ref{eq:realaxis}), performing momentum integrations, setting temperature to zero and using above identities we obtain equations satisfied by $\Gamma^{\pm}$ 
\begin{align}\label{eq:gammaplus}
&\begin{aligned}
\Gamma^+(\varepsilon_1,\varepsilon_2) =& \chi^+(\varepsilon_1\ns-\ns\varepsilon_2)& 	 						\\
-\int_{-\omega}^{0}  d\varepsilon  \Big[
&\Gamma^+(\varepsilon_1,\varepsilon)\; S(\varepsilon) \;\sign(\varepsilon\ns-\ns\varepsilon_2)\chi^-(\varepsilon\ns-\ns\varepsilon_2)    \\
+&\sign(\varepsilon_1\ns-\ns\varepsilon)\Gamma^-(\varepsilon_1,\varepsilon)\; S(\varepsilon)\; \chi^+(\varepsilon\ns-\ns\varepsilon_2)   \\
+&\Gamma^+(\varepsilon_1,\varepsilon) \;S(\varepsilon)\; \chi^+(\varepsilon\ns-\ns\varepsilon_2)  						\\
-&\Gamma^-(\varepsilon_1,\varepsilon)\; S(\varepsilon)\; \chi^-(\varepsilon\ns-\ns\varepsilon_2)  \Big]\qquad 
\end{aligned} \notag\\
&\begin{aligned}
&\Gamma^-(\varepsilon_1,\varepsilon_2) = \chi^-(\varepsilon_1\ns-\ns\varepsilon_2) \\
&\quad-2\sign(\varepsilon_1\ns-\ns\varepsilon_2)\int_{\min\varepsilon_1,\varepsilon_2}^{\max\varepsilon_1,\varepsilon_2} d\varepsilon \Gamma^-(\varepsilon_1,\varepsilon) S(\varepsilon) \chi^-(\varepsilon\ns-\ns\varepsilon_2)
\end{aligned} 
\end{align}
where $\chi^{\pm}= \nu\gamma^2(\chi^{R}\pm\chi^{A})/2$. With this definition,  $\chi^{\pm}$ absorbs the factor $\gamma^2$ which accompanies each wiggly line in the diagram as well as the factor of density of states which appears due to momentum summation in the particle-hole section, see Eq.~(\ref{eq:RAsection}). Note that both $\Gamma^{\pm}$ and $\chi^{\pm}$ are dimensionless while  $S(\varepsilon)$ has dimension of inverse energy. We suppress $\omega,q$ arguments in all functions in Eq.~(\ref{eq:gammaplus}). Here and below we omit index R in the dynamic particle-hole section $S^R(\varepsilon;\omega,q)$. 

Finally, we decouple the two equations, Eq.~(\ref{eq:gammaplus}) in terms of  functions 
$K^{\pm}$ defined as  
\begin{align}
K^+(\varepsilon_1,\varepsilon_2) =& \Gamma^+(\varepsilon_1,\varepsilon_2) + \sign(\varepsilon_1\ns-\ns\varepsilon_2) \Gamma^-(\varepsilon_1,\varepsilon_2), \notag\\
K^-(\varepsilon_1,\varepsilon_2)  =& \sign(\varepsilon_1\ns-\ns\varepsilon_2) \Gamma^-(\varepsilon_1,\varepsilon_2) \, ,
\end{align}
To describe the local fluctuation we introduce functions $X^{\pm}$~:
\begin{align}
&X^+(\varepsilon_1\ns-\ns\varepsilon_2) = \chi^+(\varepsilon_1\ns-\ns\varepsilon_2) +  \sign(\varepsilon_1\ns-\ns\varepsilon_2) \chi^-(\varepsilon_1,\varepsilon_2) \notag\\
&\qquad= -g\frac2{\pi} \ln\frac{\omega_0}{|\varepsilon_1-\varepsilon_2|} -ig \,,\notag\\
& X^-(\varepsilon_1\ns-\ns\varepsilon_2)= \sign(\varepsilon_1\ns-\ns\varepsilon_2) \chi^-(\varepsilon_1,\varepsilon_2)  \quad=  -ig \, ,
\label{eq:newvars}
\end{align}
where we have used a form of a singular correlation, Eq.~(\ref{eq:local}), valid at $|\varepsilon_1-\varepsilon_2|<\omega_0$.
Equations for $K^{\pm}$ have the form~:
\begin{align}
&K^+(\varepsilon_1,\varepsilon_2) =\!\! X^+(\varepsilon_1\ns-\ns\varepsilon_2) -\!\!\!\int_{-\omega}^{0} \!\!\!\!\!d\varepsilon K^+(\varepsilon_1,\varepsilon)\; S(\varepsilon) \; X^+(\varepsilon\ns-\ns\varepsilon_2) 
\label{eq:decoupled-plus}\\
&K^-(\varepsilon_1,\varepsilon_2) =\!\! X^-(\varepsilon_1\ns-\ns\varepsilon_2) - \!\!2\!\!\!\!\!\!\!\!\!\!\int\limits_{\min(\varepsilon_1,\varepsilon_2)}^{\max(\varepsilon_1,\varepsilon_2)} \!\!\!\!\!\!\!\!\!\!d\varepsilon 
K^-(\varepsilon_1,\varepsilon) S(\varepsilon) X^-(\varepsilon\ns-\ns\varepsilon_2)
\label{eq:decoupled-minus}
\end{align}
The second equation follows immediately from the second equation in Eq.~(\ref{eq:gammaplus}) after multiplying both sides by $\sign(\varepsilon_1-\varepsilon_2)$ and observing that under the integral $\sign(\varepsilon_1-\varepsilon)\sign(\varepsilon-\varepsilon_2)$ is always positive (and equal $1$). To obtain Eq.~(\ref{eq:decoupled-plus}) we observe that expression under the integral in the first equation in Eq.~(\ref{eq:gammaplus}) can be rewritten as $K^+SX^+ - \Gamma^-S\chi^- - K^-SX^-$ (we suppressed all function arguments). Let us look closely at the last two terms here,  ${\scriptsize\int_{-\omega}^0[-\!1\!\!-\!\sign(\varepsilon_1\!\!-\!\!\varepsilon)\sign(\varepsilon-\varepsilon_2)] \Gamma^-S\chi^-}$. The square bracket vanishes for $\varepsilon$ outside the interval $\varepsilon_1,\varepsilon_2$, so that effectively the limits of the integral are $[{\min(\varepsilon_1,\varepsilon_2)}..{\max(\varepsilon_1,\varepsilon_2)}]$. For $\varepsilon$ within this interval the square bracket equals $-2$. Therefore the last two terms amount to ${\scriptsize-2\int_{\min(\varepsilon_1,\varepsilon_2)}^{\max(\varepsilon_1,\varepsilon_2)} \Gamma^-S\chi^-}$. In the equation for $K^+$ this expression cancels identically the integral term [multiplied by $\sign(\varepsilon_1-\varepsilon_2)$] in Eq.~(\ref{eq:gammaplus}). 

\section{Density-density correlation function in MFL.}

We first demonstrate that the singular vertex determined by Eq.~(\ref{eq:gammaplus}) ensures  particle conservation. We calculate the density correlation function and check that 
\begin{align}\label{eq:rhorhotot}
\lim\limits_{q\rightarrow0}\av{\rho\rho}^{\text{tot}}_{\omega,q}=0
\end{align}
As discussed above, in terms of dynamic part of the correlation function, $\av{\rho\rho}_{\omega,q}^{\text{dyn}}$,  this condition takes the form  $\lim\limits_{q\rightarrow0}\av{\rho\rho}_{\omega,q}^{\text{dyn}}=\text{const}$. We sum the series in Eq.~(\ref{eq:density}) and obtain
\begin{align}\label{eq:densitydensity}
&\av{\rho\rho}_{\omega,q}^{\text{dyn}}  = 
2\nu(1+B_0)^2 \frac{ Z(\omega,q) }{ 1 - B_0Z(\omega,q) }
\end{align}
where the factor 2 is due to spin summation. The factor $(1+B_0)^2$ comes from the external vertices (``triangles'') in Eq.~(\ref{eq:density}); these include only non-singular renormalizations. The function $Z(\omega,q)$ represents the ``dotted'' section, Eq.~(\ref{eq:dressedsection}), summed over fermion frequency,  
\begin{align}\label{eq:defZ}
&Z(\omega,q) = -\int_{-\omega}^0d\varepsilon S(\varepsilon) \notag\\
&\qquad +\int_{-\omega}^0 d\varepsilon_1d\varepsilon_2 \; S(\varepsilon_1) K^+(\varepsilon_1,\varepsilon_2) S(\varepsilon_2) \,.
\end{align}
We suppressed $\omega,q$ arguments on the right hand side. Note that vertex $\Gamma_s$ is represented here by the function  $K^+$ only. 

Let us make the following comment. Ignoring contribution from the branch cut at $\Im(\varepsilon_1-\varepsilon_2)=0$  roughly corresponds to ignoring  $\Gamma^{-}$ and $\chi^{-}$ everywhere in Eq.~(\ref{eq:decoupled-plus}), i.e., we replace $K^+$ and $X^+$ with $\Gamma^+$ and $\chi^+$ respectively. We obtain a ladder equation $\Gamma^+ =  \chi^{+} - \omega \Gamma^+ S \chi^{+}$ (where $S(\omega,q)$ is given by Eq.~(\ref{eq:RAsection}) and one has to set the self-energy to zero for consistency). This equation introduces static (logarithmic) renormalizations due to rescattering by $\chi^{+}$ and leads to density correlation function that does not satisfy particle conservation condition. Proper solution of Eq.~(\ref{eq:decoupled-plus}) cures both problems, particle conservation is restored and logarithmic renormalizations disappear. This supports the statement made in Sec.~I that in consistent microscopic description of MFL one has to consider the dynamic effect of local fluctuation in the vertex $\Gamma_s$ when their effect has been accounted for in fermion self-energy.

Physically sensible approximation to solution of Eq.~(\ref{eq:decoupled-plus}), i.e., the one that ensures particle conservation, can be devised as follows. First observe that in the interval $(-\omega<\varepsilon<0)$ (i) the imaginary part of $\Sigma^R(\varepsilon+\omega)-\Sigma^A(\varepsilon)$ is independent of $\varepsilon$ for marginal self-energy and (ii) its real part varies only weakly. Point (ii) follows because $\Re\Sigma(\varepsilon)$ differs from linear function (for which the statement would  hold exactly) only  by a logarithmic factor. Therefore to logarithmic accuracy we can approximate $r(\varepsilon;\omega) \equiv  \Sigma^R(\varepsilon+\omega)-\Sigma^A(\varepsilon)$ in the interval  $-\omega<\varepsilon<0$ by a function that does not depend on $\varepsilon$. To be specific, we define 
\begin{align}
r(\omega) \equiv&  [\Sigma^R(\varepsilon+\omega)-\Sigma^A(\varepsilon)]_{\varepsilon=-\omega/2}\notag\\
=&  -i g \omega - \frac{2g}{\pi} \left( \ln\frac{2\omega_0}{|\omega|} +1 \right)\omega
\end{align}
Within this approximation the weak $\varepsilon$-dependence of the dynamic particle-hole section $S(\varepsilon;\omega,q)$  in the interval  $-\omega<\varepsilon<0$ is ignored, i.e., we replace $S(\varepsilon;\omega,q)$ with
\begin{align}\label{eq:Somegaq}
S_{\omega,q} = \int\frac{d\theta}{2\pi} \; \frac{1}{\omega-qv_F\cos\theta-r(\omega)}
\end{align}
everywhere. With this, $\varepsilon$-integration can be performed in Eqs.~(\ref{eq:decoupled-plus}) and Eq.~(\ref{eq:defZ}); one has to use an identity $\int_{-\omega}^0 d\varepsilon X^+(\varepsilon-\varepsilon') =\Sigma^R(\omega+\varepsilon') -\Sigma^A(\varepsilon')$ which is satisfied for marginal Fermi liquid self-energy, Eqs.~(\ref{eq:selfenergy}) and~(\ref{eq:selfenergy-real}) in an interval $-\omega<\varepsilon'<0$. Finally, Eq.~(\ref{eq:decoupled-plus}) reduces to an algebraic equation
\begin{align}\label{eq:rhorho}
&\tilde{K}_{\omega,q}^+ \approx \omega r(\omega) - S_{\omega,q} r(\omega)  \tilde{K}_{\omega,q}^+  
\end{align}
where $\tilde{K}_{\omega,q}^+=\int_{-\omega}^0 d\varepsilon_1d\varepsilon_2  K_{\omega,q}^+(\varepsilon_1,\varepsilon_2)$. Using solution of this equation in Eq.~(\ref{eq:defZ}) we obtain~:
\begin{align}\label{eq:Z}
&Z(\omega,q) \approx  -\omega S_{\omega,q}+ S_{\omega,q}^2 \tilde{K}_{\omega,q}^+  \notag\\
&\qquad=- \frac{\omega S_{\omega,q}}{1+S_{\omega,q}r(\omega)} \,.
\end{align}
To check the particle conservation condition Eq.~(\ref{eq:rhorhotot}) we have to take the limit $q=0$ in Eq.~(\ref{eq:densitydensity}). In this limit the particle-hole section $S(\omega,q\rightarrow0)\equiv S_0(\omega)=1/[\omega-r(\omega)]$. Combining Eqs.~(\ref{eq:densitydensity})  and (\ref{eq:Z}) we find 
\begin{align}\label{eq:aaa}
\lim\limits_{q\rightarrow0}\av{\rho\rho}_{\omega,q}^{\text{dyn}}=-2\nu(1+B_0)\,. 
\end{align}
This form of the density correlation function ensures particle conservation because the right-hand side is real $\omega$-independent constant. 

Using  $\av{\rho\rho}^{\text{tot}}=\av{\rho\rho}^{\text{dyn}}+\av{\rho\rho}^{\text{static}}$ we find  $\av{\rho\rho}_{\omega,q}^{\text{static}}=2\nu(1+B_0)$. Let us comment on this result. We find that in MFL the compressibility acquires only non-singular renormalizations,  $dn/d\mu=2\nu/(1+F_0^s)$,  where $\nu$ is the quasiparticle density of states and where the non-singular Fermi liquid amplitude $F_0^s$ is related to $B_0$ via $1+F_0^s = 1/(1+B_0)$. This is to be contrasted with the thermodynamic density of states that controls specific heat, $\gamma=2\nu/z$,  which is enhanced by logarithmic factor in $1/z=1-d\Sigma/d\varepsilon$. Let us note that singular dynamics of quasiparticles near the Fermi surface cannot affect compressibility $dn/d\mu$ on general grounds\cite{SovSciRev1983}. Indeed, the shift of the chemical potential only leads to a shift of the Fermi energy and with it shift in all the physics that is ``pinned'' to it. This leaves derivative  $dn/d\mu$ free of any singularity, see also discussion in Ref.~\onlinecite{Kotliar1991}.

Since we now have conserving approximation for a particle-hole ladder, we can calculate conductivity using equation of continuity 
\begin{align}\label{eq:aaa}
\sigma(\omega) = \lim\limits_{q\rightarrow0} \frac{\omega}{q^2} \Im \av{\rho\rho}_{\omega,q}^{\text{tot}} \,.
\end{align}
Unlike previous discussion, we now have to keep the $q$-dependence in $S_{\omega,q}$ to lowest order in $q$. It follows from Eq.~(\ref{eq:Somegaq}) that
\begin{align}\label{eq:smallq}
&S(qv_F\ll\omega) \approx \frac{1}{\omega-r(\omega)+id(\omega)q^2}
\end{align}
where we have introduced  $d(\omega) = i[\omega-r(\omega)]^{-1} v_F^2/2$, see Ref.~\onlinecite{notation}. Substituting this in $\av{\rho\rho}^{\text{tot}}=\av{\rho\rho}^{\text{dyn}}+\av{\rho\rho}^{\text{static}}$ and using Eq.~(\ref{eq:rhorho}) we obtain
\begin{align}
&\av{\rho\rho}_{\omega,q}^{\text{tot}}|_{qv_F\ll\omega} \approx 2\nu(1+B_0)^2 \frac{d(\omega) q^2 }{-i\omega(1+B_0) + d(\omega) q^2} \,.
\end{align}
We find
\begin{align}
&\sigma(\omega) = 2\nu(1+B_0) \Re d(\omega) \notag\\
&\qquad =2\nu(1+B_0)\frac{ v_F^2}2 \Im\frac{-1}{\omega -r(\omega)}
\end{align}
This is equivalent to the result obtained with direct calculation of conductivity bubble \cite{Varma, Abrahams1996} [except for the Fermi-liquid factor $(1+B_0)$] where vertex corrections  were argued to be zero because they are momentum independent while the vertex coupling to the vector-potential for calculation of conductivity is the momentum vector. The results here are a formal justification of such simplifications. 

\section{Conclusions.}

In conclusion, we have demonstrated through microscopic theory that conservation laws are obeyed by a marginal Fermi liquid  as well as provided a consistent theory of the  hydrodynamic transport properties. We expect that our approach is useful for other problems \cite{VNW} with singular low energy properties of interacting fermions.
\begin{acknowledgments}
We thank A.M.~Finkel'stein for stimulating discussions. A.S. would like to acknowledge hospitality of Aspen Center for Physics where part of this work was done. 
\end{acknowledgments}


\begin{thebibliography}{99}
\bibitem{rotor} In this paper we assume $\gamma$ to be constant on the Fermi surface. Recently the form of the coupling constant has been found, see Ref.~\onlinecite{ASV}. The angular dependence of the coupling constant has the form, $\gamma(kk')\propto i k\times k'$. In the particle-hole channel its angular decomposition contains only even harmonics. For instance, using the above form in the singular vertex, Eq.~(\ref{eq:elementaryvertex}), we obtain $(i k\times k')^2 \propto 1-cos(2\theta-2\theta')$. Only the zero harmonic is important in the analysis of density correlation function.
\bibitem{anotherBC} There is another branch cut at  $\Im\varepsilon_1+\Im\varepsilon_2=0$  which follows from the absorptive part in the Cooper channel. This other channel will not be considered here.
\bibitem{notation} This notation is motivated by apparent similarity with diffusive system where $d(\omega)=v_f^2\tau/2=D$ because $S_0\approx-i\tau$ when $\omega,qv_F\ll\tau^{-1}$. The similarity does not go beyond notation: in marginal Fermi liquid the motion of quasiparticles is not diffusive because Eq.~(\ref{eq:smallq}) is valid only at ultra-small momenta $qv_F\ll\omega$.
\end{thebibliography}
\end{document}